\documentclass[12pt,preprint]{aastex}
\pdfoutput=1

\newcommand{\eqref}[1]{(\ref{#1})}

\shorttitle{Gravitational Instability and Star Formation}
\shortauthors{Yang et al.}

\begin{document}

\title{Large-Scale Gravitational Instability and Star Formation\\
       in the Large Magellanic Cloud}

\author{Chao-Chin Yang, Robert A. Gruendl, and You-Hua Chu}
\affil{Department of Astronomy, University of Illinois, Urbana, IL~61801}
\email{cyang8@astro.uiuc.edu, gruendl@astro.uiuc.edu, chu@astro.uiuc.edu}

\and

\author{Mordecai-Mark Mac Low}
\affil{Department of Astrophysics, American Museum of Natural History,
       New York, NY~10024-5192}
\email{mordecai@amnh.org}

\and

\author{Yasuo Fukui}
\affil{Department of Physics and Astrophysics, Nagoya University,
       Chikusa-ku, Nagoya~464-8602, Japan}
\email{fukui@a.phys.nagoya-u.ac.jp}

\begin{abstract}
Large-scale star formation in disk galaxies is hypothesized to be driven by
global gravitational instability.
The observed gas surface density is commonly used to compute the strength of
gravitational instability, but according to this criterion star formation often
appears to occur in gravitationally stable regions.
One possible reason is that the stellar contribution to the instability has
been neglected.
We have examined the gravitational instability of the Large Magellanic Cloud
(LMC) considering the gas alone, and considering the combination of collisional
gas and collisionless stars.
We compare the gravitationally unstable regions with the on-going star
formation revealed by \emph{Spitzer} observations of young stellar objects.
Although only 62\% of the massive young stellar object candidates are in
regions where the gas alone is unstable, some 85\% lie in regions unstable due
to the combination of gas and stars.
The combined stability analysis better describes where star formation occurs.
In agreement with other observations and numerical models, a small fraction of
the star formation occurs in regions with gravitational stability parameter
$Q > 1$.
We further measure the dependence of the star formation timescale on the
strength of gravitational instability, and quantitatively compare it to the
exponential dependence expected from numerical simulations.
\end{abstract}


\section{INTRODUCTION}

Observations of nearby galaxies suggest that there exists a gas surface density
threshold for star formation in a galactic disk \citep{rK89,MK01}.
This threshold is hypothesized to be formed by the threshold for 
gravitational stability of the disk, although alternative explanations
have been offered \citep[e.g.,][]{jS04}.
Gravitationally unstable regions are commonly identified by the Toomre
criterion \citep{aT64,GL65}, which considers the stability of a
single-component disk against axisymmetric perturbations.
However, this criterion, when applied to observed gas surface
densities, does not always 
correctly predict the locations of star formation \citep[e.g.,][]{MK01,WB02}.
Stellar contribution to the gravitational instability could be important;
for example, the giant molecular clouds in the Large Magellanic Cloud (LMC)
have been shown to correlate with stellar surface density, indicating the
increase of star formation towards regions of higher stellar gravity
\citep{yF07}.

To crudely include the stellar contribution, \citet{JS84} derived the stability
criterion for a two-fluid disk.
This stability criterion differs from the Toomre criterion by only a linear
correction factor, when the gas-to-stellar ratios of surface density and
velocity dispersion are constant across a galaxy, or when the gas and stellar
components are only weakly coupled \citep{WS94}.
To correctly include the stellar contribution, \citet{cG92} and
\citet{rR01} analyzed the 
stability of a composite disk consisting of a collisional gas component and a
population of collisionless stellar components.
The resulting instability criterion has been used to explain the dust
morphology of edge-on disk galaxies \citep*{DY04}.

We study the relationship between gravitational stability and star
formation in the LMC.
The LMC is chosen because high linear resolution can be obtained
\citep[1\arcsec\ corresponds to 0.25~pc at 50~kpc,][]{mF99}, its internal and
foreground extinctions are small, and its moderate inclination allows
spatial and kinematical mapping of the disk with minimum confusion along the
line of sight.
Both stars and the interstellar medium (ISM) of the LMC have been extensively
surveyed.
Recent \emph{Spitzer Space Telescope} observations have been used to identify
massive young stellar objects (YSOs), which mark the sites of current star
formation (R.~A. Gruendl et al., in preparation; B.~A. Whitney et al.,
in preparation).
Massive YSOs provide better probes than \ion{H}{2} regions to study the
relationship between the ISM and star formation because their stellar energy
feedback has not significantly altered the distributions and physical
conditions of the ambient ISM.

In this paper, we investigate the relationship between global star formation
and gravitational instability in the LMC using two different stability
analyses: assuming that the gaseous disk is decoupled from the stellar disk and
considering only the gravity of the gas (\S~\ref{S:Q_g}), or assuming a
collisionless stellar disk and a collisional gas disk and using the
\citet{rR01} stability criterion (\S~\ref{S:Qsg}).
The results are discussed in \S~\ref{S:disc}.

\section{GRAVITATIONAL INSTABILITY} \label{S:GI}

\subsection{Gas Alone} \label{S:Q_g}
A thin, differentially rotating gaseous disk is unstable against
axisymmetric perturbations when
\begin{equation} \label{E:Q_g}
  Q_g \equiv \frac{\kappa c_g}{\pi G\Sigma_g} < 1,
\end{equation}
where $Q_g$ is generally referred to as the Toomre $Q$ parameter for a
gaseous disk, $\kappa$ is the epicyclic frequency, $c_g$ is the isothermal
sound speed of the gas, $G$ is the gravitational constant, and
$\Sigma_g$ is the unperturbed surface density of the gas \citep{GL65}.

To apply this analysis to a disk galaxy, approximations need to be made.
First, rather than consisting of a single isothermal component, the real ISM
consists of gas with temperatures ranging from 10~K to $10^7$~K in multiple
phases including cold atomic and molecular, warm atomic and ionized, and hot
ionized gas.
However, the gas mass is dominated by cold atomic and molecular material in
turbulent motion, except possibly in star-forming regions where much of the ISM
is ionized.
Therefore, we consider only the neutral atomic and molecular gas mass.
The atomic medium of galaxies generally has nearly constant velocity dispersion
\citep*{VS82,SV84,DH90,PR07}, and has a vertical thickness that is well modeled
by assuming hydrostatic equilibrium with the gravitational potential at the
velocity dispersion \citep[e.g.,][]{sM95}.
We assume the gas disk of the LMC has an effective sound speed representative
of the velocity dispersion $c_g = 5$~km~s$^{-1}$ \citep*{DB06,PR07}.
Second, $\Sigma_g$ in equation~\eqref{E:Q_g} is the \emph{unperturbed} gas
surface density, while the observed gas surface density is \emph{perturbed}.
The surface density of a gravitationally unstable region grows exponentially as
it collapses; thus, using the observed surface density in
equation~\eqref{E:Q_g} will correctly diagnose gravitationally unstable
regions, but $Q_g$ may be underestimated if collapse has set in.
Therefore, we adopt the observed gas surface density for $\Sigma_g$ in
equation~\eqref{E:Q_g} to identify gravitationally unstable regions.

The neutral atomic and molecular components of the ISM in the LMC have been
well surveyed.
For the neutral atomic component, we use the \ion{H}{1} column density map
derived from the combined data set of the Australia Telescope Compact Array and
the Parkes multibeam receiver \citep{KS03}.
This \ion{H}{1} map has a resolution of $\sim$1\arcmin, or 15~pc.
For the molecular component, we use the NANTEN CO survey of the LMC conducted
by \citet{FM99,FM01}.
These maps have a resolution of $\sim$2\farcm6, or 40~pc.
We scale maps of integrated column density by the CO-to-H$_2$ conversion
factor, $X = 5.4\times10^{20}$~H$_2$ atoms cm$^{-2}$ (K km s$^{-1}$)$^{-1}$
\citep{BF07} to obtain the H$_2$ mass distribution.
Both the \ion{H}{1} and H$_2$ maps are clipped at the 1$\sigma$ level
to prevent negative densities and minimize the effects of noise.
Furthermore, we account for He and heavier elements by assuming that
the mass fraction of H is $X \sim 0.7$.
From the \ion{H}{1} map, we find a total mass for the neutral atomic component
of $5.1\times10^8~M_\sun$, consistent with the value found by \citet{KS98}.
For the molecular component, we find a total mass of $6.6\times10^7~M_\sun$,
which is on the high side of the 4--7$\times10^7~M_\sun$ given by
\citet{nM99,nM01}, as we used a higher CO-to-H$_2$ conversion factor and
included He and heavy elements.
We add these two maps to obtain a map of total gas surface density with a
resolution of 40~pc pixel$^{-1}$.
The resulting map is shown in Figure~\ref{Fi:maps}a.

The remaining parameter needed in equation~\eqref{E:Q_g} is the
epicyclic frequency $\kappa$, defined by
\begin{equation}
  \kappa^2 \equiv \frac{2V^2}{R^2}\left(1 + \frac{R}{V}\frac{dV}{dR}\right),
\end{equation}
where $R$ is the galactocentric distance and $V = V(R)$ is the circular
velocity as a function of $R$.
To calculate $\kappa$ at a given $R$ in the LMC, we use the rotation curve
derived by \citet{KS98}.
The rotation curve is a best-fit to the \ion{H}{1} and the carbon star
\citep{KD97} measurements inside and outside of about 3.2~kpc, respectively.
We find the derivatives of the circular velocity and thus the epicyclic
frequency by central finite differences with an increment of 250~pc, then we
use a cubic spline to interpolate the discrete results to a continuous domain.
Finally, the center and the orientation of the disk of the LMC need to be
specified.
To be consistent with the \ion{H}{1} kinematics, we adopt the same center and
orientation determined by \citet{KS98}.
The kinematic center of the LMC disk is at $\alpha = 05^{\rm h}$17\fm6 and
$\delta = -69$\arcdeg02\arcmin\ (J2000) with an inclination of $i = 33\degr$
and a line of nodes at a P.A. of $-12\degr$.
Using this geometry, the total gas surface density map is deprojected to
determine the galactocentric distance of each pixel, allowing us to construct a
map of $Q_g$ calculated from equation~\eqref{E:Q_g}.
This map is shown in Figure~\ref{Fi:maps}b.

The contours in Figure~\ref{Fi:maps}b delineate the critical boundary
$Q_g = 1$ that surrounds regions gravitationally unstable due to gas alone. 
Also marked in the figure are candidate massive YSOs from R.~A. Gruendl et al.
(in preparation), indicating regions of on-going star formation.
In the top panel of Figure~\ref{Fi:hist}, we show the number distribution of
YSO candidates and the corresponding cumulative fraction with respect to $Q_g$.
While $\sim$62\% (153 of 245) of the YSO candidates do lie in regions with
$Q_g < 1$, the rest are distributed in regions with values as high as $Q_g \sim
3$.

Some of the candidates with $Q_g < 1$ might be argued not to result from
large-scale gravitational instabilities because the unstable regions in which
they reside are too small.
If we exclude regions $\la 100$~pc in diameter, only 47\% (116) of the YSO
candidates would be associated with gravitationally unstable regions by this
criterion.
A frequently used empirical correction for the contributions from the stellar
component \citep{JS84,WS94} and the effect of the disk scale height
\citep{JS84} is to raise the critical value to $Q_g \sim 1.4$
\citep{rK89,MK01}.
Even if we apply this linear correction factor to the Toomre criterion,
$\sim$25~\% of the YSO candidates still appear to reside in stable regions.
Considering the stability of only the gas disk or applying a simple linear
correction to the Toomre criterion may not be sufficient to account for star
formation activity.

\placefigure{Fi:hist}

\subsection{Gas and Stars Together} \label{S:Qsg}

We now also consider the contribution of stars to gravitational instability.
To include them, we follow Rafikov's (2001) treatment of a disk galaxy
consisting of a collisional gas disk and a collisionless stellar disk.
The instability condition becomes
\begin{equation} \label{E:Qsg}
  \frac{1}{Q_{sg}}
  \equiv \frac{2}{Q_s}\frac{1}{q}\left[1 - e^{-q^2}I_0\left(q^2\right)\right] +
         \frac{2}{Q_g}R\frac{q}{1 + q^2 R^2}
  > 1.
\end{equation}
In equation~\eqref{E:Qsg}, $Q_g$ is the stability parameter for the gas derived
by \citet{GL65}, as defined in equation~\eqref{E:Q_g}, while
\begin{equation}
   Q_s \equiv \frac{\kappa\sigma_s}{\pi G\Sigma_s}
\end{equation}
is the stability parameter for the stars derived by \citet{aT64}, where
$\sigma_s$ is the stellar radial velocity dispersion, $\Sigma_s$ is the stellar
surface density;
$R \equiv c_g / \sigma_s$ and $q \equiv k\sigma_s / \kappa$ are two
dimensionless parameters, with $k$ being the wavenumber of the axisymmetric
perturbations;
and $I_0$ is the Bessel function of order zero.

Similarly to our procedure in \S~\ref{S:Q_g}, we use local observed
values to evaluate $Q_{sg}$. 
To estimate the stellar surface density $\Sigma_s$, we use the number density
of red giant branch (RGB) and asymptotic giant branch (AGB) stars, as they are
part of the old stellar population and therefore trace the overall mass
distribution of the stellar disk.
RGB and AGB stars are luminous and distinctive in the $[J - K_s]$ vs.\ $K_s$
color-magnitude diagram and much less confused by faint background galaxies or
Galactic stars.
To select these stars, we follow a procedure similar to that outlined by
\citet{rV01} but use only the Two Micron All Sky Survey Point Source Catalogue
\citep{mS06} and the criteria: $[J - K_s] < (22 - [K_s]) / 10.5$ and
$[K_s] < 14.5$.  
The source counts are then binned with a resolution of 40~pc pixel$^{-1}$ and
Gaussian smoothed with a dispersion of 100~pc.
The smoothing scale is chosen to minimize pixel-to-pixel variations due to the
small number density of the tracer RGB and AGB stars, while retaining good
spatial resolution for the analysis; this scale is also small compared to the
large-scale gravitational instability in the analysis.
In the resulting map, a faint Galactic background with a gradient is present.
Therefore, we fit and subtract this background with exponential and linear
dependencies on the Galactic latitude and longitude, respectively.
Finally, by adopting the total stellar mass of $2 \times 10^9~M_\sun$
estimated by \cite{KS98}, we normalize this map to obtain the absolute stellar
surface density map shown in Figure~\ref{Fi:maps}c
\citep[cf. Fig.~2 of][]{rV01}.
We adopt a constant stellar radial velocity dispersion of 15~km~s$^{-1}$, after
considering the velocity dispersions of carbon stars
\citep{KD97,AN00,GG00,VA02}, red supergiants \citep{PM89,OM07}, and young
globular clusters \citep{FI83}.

With these values assessed, the parameter $Q_{sg}$ defined in
equation~\eqref{E:Qsg} now depends only on the perturbation wavenumber $k$, or
equivalently, the perturbation wavelength $\lambda = 2\pi / k$ at 
each pixel (see Fig.~\ref{Fi:qsgwl}).
Since $Q_{sg}$ decreases with the degree of gravitational instability and our
objective is to locate gravitationally unstable regions, we define its value at
each pixel as its global minimum with respect to $\lambda$.
The derived $Q_{sg}$ map of the LMC is shown in Figure~\ref{Fi:maps}d, with the
identified YSOs overlaid.

Each point is unstable to perturbations over a finite range of wavelengths, as
shown by Figure~\ref{Fi:qsgwl}.
If the size of a gravitationally unstable region is smaller than the minimum
unstable wavelength $\lambda_\mathrm{min}$, no perturbation inside the region
can have long enough wavelength to drive an instability.
Therefore, we also find $\lambda_\mathrm{min}$ in each pixel, and compare this
value with the size of each region containing YSO candidates by visual
inspection.
The contours of $Q_{sg} = 1$ are plotted over grayscale presentations of
$\lambda_\mathrm{min}$ in Figure~\ref{Fi:lmin} so that the sizes of the
$Q_{sg} < 1$ regions can be compared with the minimum unstable wavelengths
directly.

\placefigure{Fi:qsgwl}

Figure~\ref{Fi:maps}d shows that, because of the stellar contribution, a much
larger fraction of the galaxy is unstable than would have been deduced just
from the gas contribution shown in Figure~\ref{Fi:maps}b.
The vast majority of the massive YSO candidates are in fact located within
gravitationally unstable regions, in sharp contrast to the results in
\S~\ref{S:Q_g} where only the gas is considered.
Particularly noticeable is the vicinity of the LMC bar region, where on-going
star formation is observed.
This bar region, having low gas surface densities and high epicyclic
frequencies, is stable when only the gas disk is considered, but becomes
unstable when the stellar component is added.
The bottom panel of Figure~\ref{Fi:hist} shows the number distribution of YSO
candidates and the corresponding cumulative fraction with respect to $Q_{sg}$.
About 86\% (212 of 245) of the YSO candidates are located in regions where
$Q_{sg} < 1$.
Figure~\ref{Fi:lmin} shows 12 of the 212 YSOs in probably stable regions having
$Q_{sg} < 1$ but region size $< \lambda_\mathrm{min}$.
Removing these 12 still leaves 82\% of the YSOs whose formation is attributed
to gravitational instability of the disk of stars and gas.
One caution is that finite thickness of the disk may lower the critical
$Q_{sg}$ value slightly, depending on the scale height of the disk relative to
the unstable wavelengths \citep{JS84}.
Given an estimated scale height of $\sim$180~pc \citep{KD99} and
kiloparsec-scale unstable wavelengths in the LMC, this effect may be small.
The results demonstrate that the stellar contribution to gravitational
instability can be significant.
Gravitational instability of the full disk appears responsible for most
large-scale star formation activity.

\placefigure{Fi:lmin}

\section{DISCUSSION AND CONCLUSIONS} \label{S:disc}

\begin{figure}
   \plotone{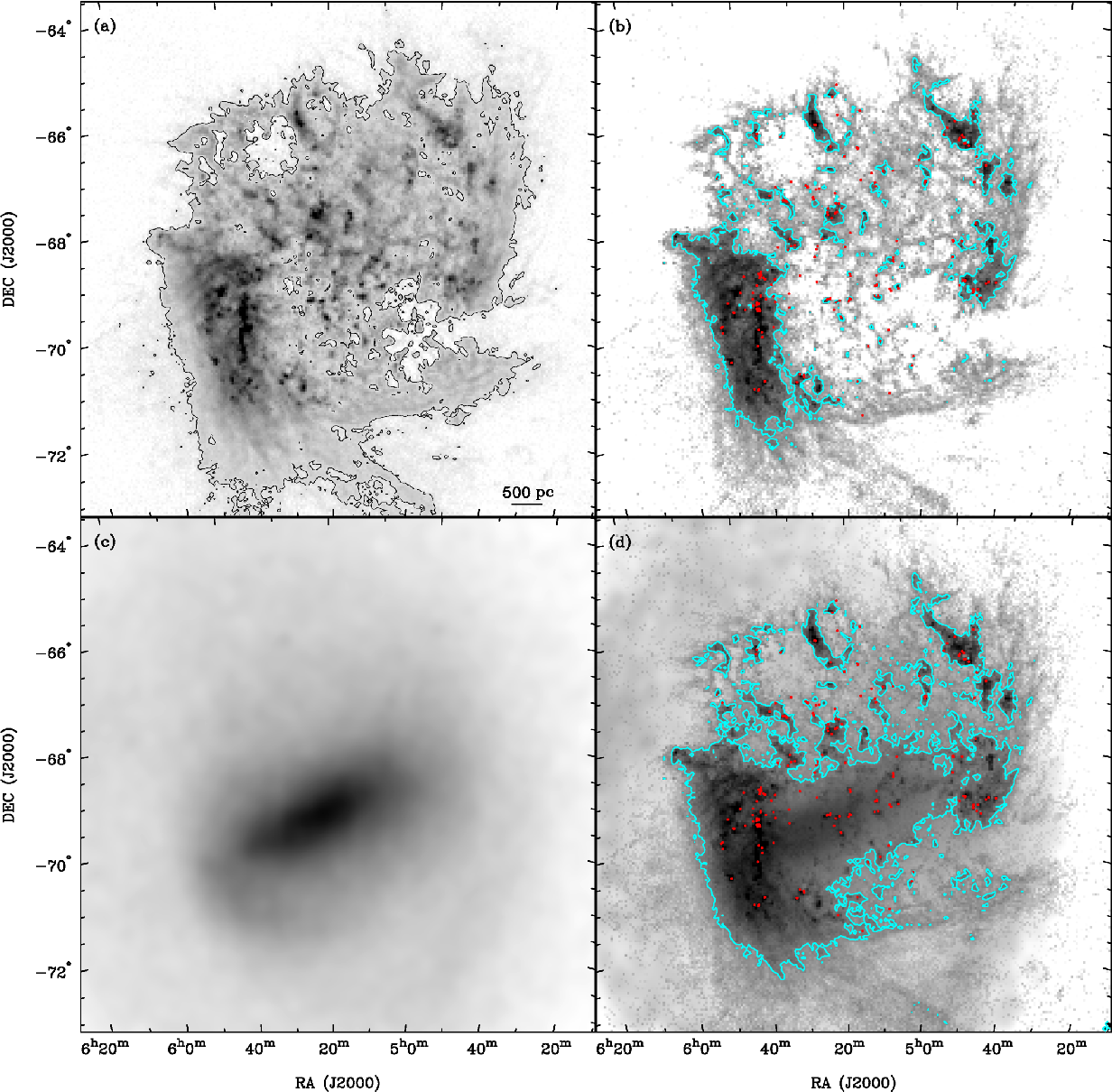}
   \caption{(a) Total gas surface density distribution in the Large Magellanic
      Cloud. Only neutral species are included in this map. The grayscale runs
      linearly from 0 to 100~$M_\sun$ pc$^{-2}$. The contour lines show a
      surface density of 4~$M_\sun$ pc$^{-2}$.
      (b) Comparison between massive star-forming sites traced by
      young stellar object candidates (\emph{red dots}) and regions
      where the gas alone is gravitationally unstable.
      The Toomre parameter for the gas $Q_g$ is shown in grayscale, running
      logarithmically from 5.0 to 0.2.
      The \emph{solid} lines delineate the critical value of $Q_g = 1$ inside
      which the regions are gravitationally unstable.
      (c) Total stellar surface density distribution.
      The grayscale runs linearly from 0 to 200~$M_\sun$ pc$^{-2}$.
      (d) As in (b), but for regions where the gas and stars together
      are gravitationally unstable, rather than just the gas alone.}
   \label{Fi:maps}
\end{figure}

\begin{figure}
   \plotone{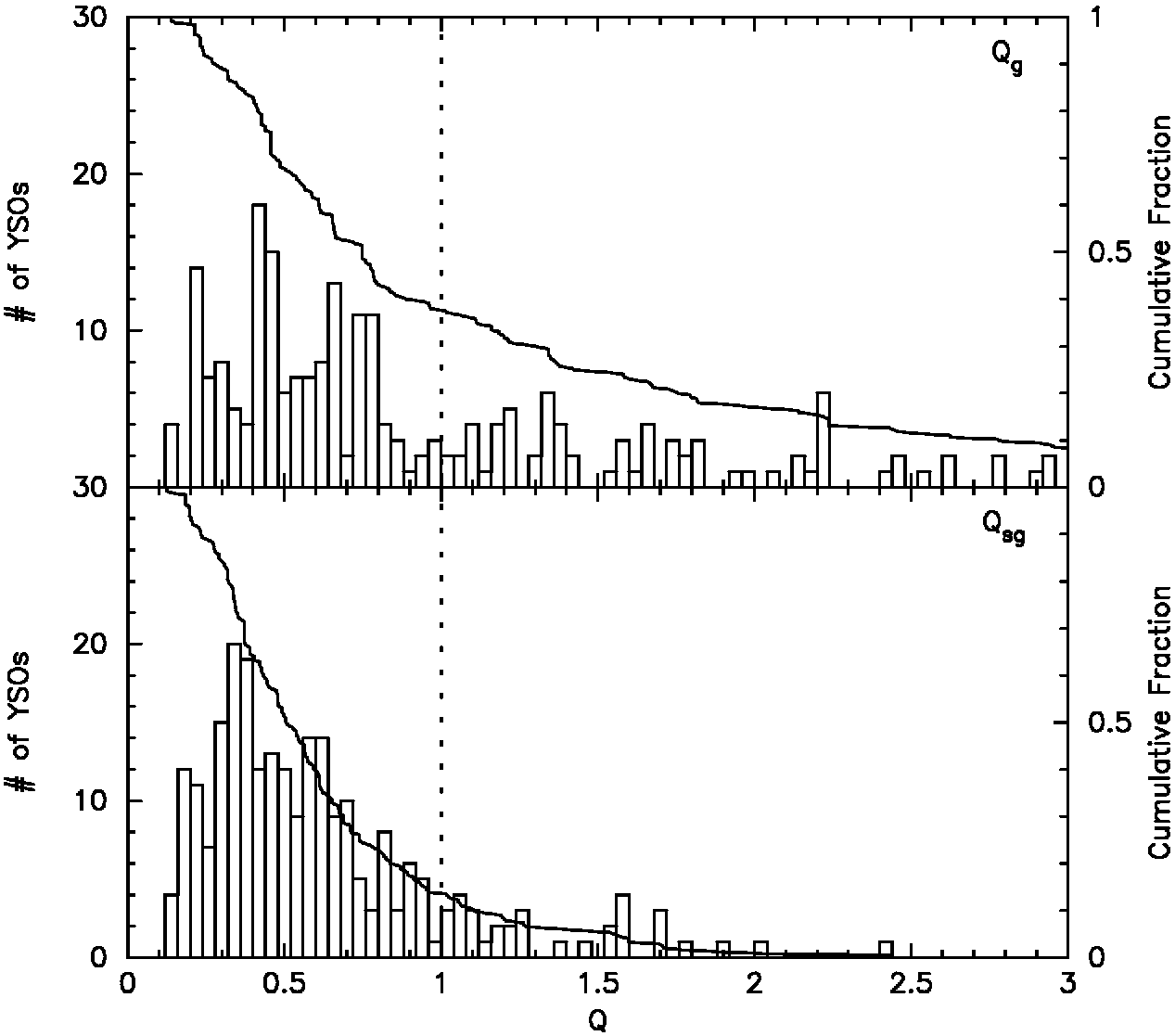}
   \caption{Number distribution of YSO candidates with respect to the Toomre
      stability parameter of the pixel in which they are located.
      The \emph{top} and \emph{bottom} panels show respectively the Toomre
      parameters for the gas alone $Q_g$, and the stars and gas together
      $Q_{sg}$.
      The \emph{solid} lines show the cumulative fractions of YSOs with
      decreasing $Q$.
      The vertical \emph{dotted} lines denote the critical value $Q = 1$.}
   \label{Fi:hist}
\end{figure}

\begin{figure}
   \plotone{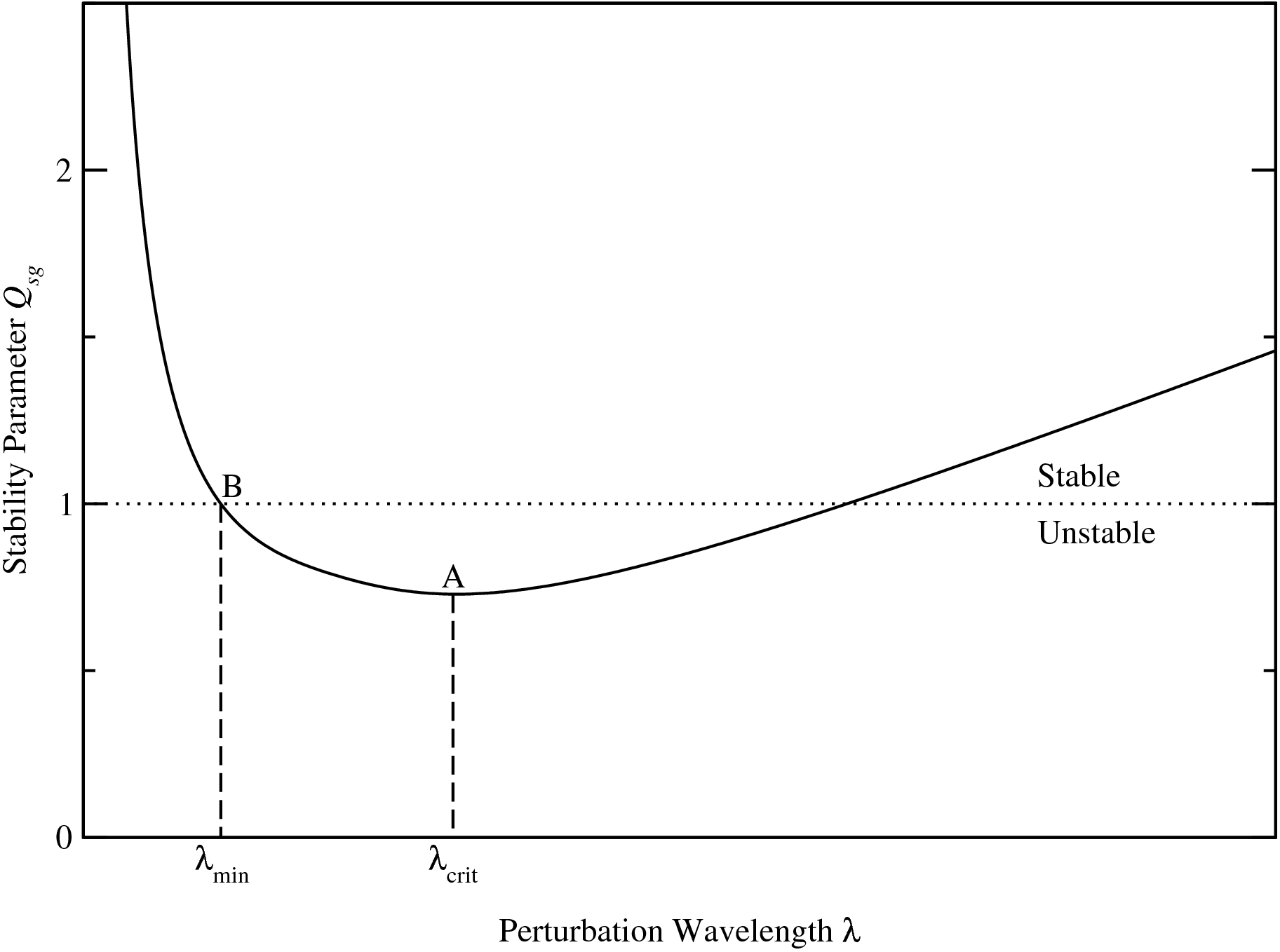}
   \caption{Schematic plot of the $Q_{sg}$ parameter as a function of
      wavelength $\lambda$ of radial perturbations.
      The disk of stars and gas is stable against perturbation of a given
      $\lambda$ when $Q_{sg} > 1$; otherwise, it is unstable.
      We measure the value of $Q_{sg}$ at its minimum in each pixel, occurring
      at $\lambda = \lambda_\mathrm{crit}$ (point A).
      We also find the minimum wavelength $\lambda_\mathrm{min}$ such that the
      disk is marginally unstable (point B).}
   \label{Fi:qsgwl}
\end{figure}

\begin{figure}
   \plotone{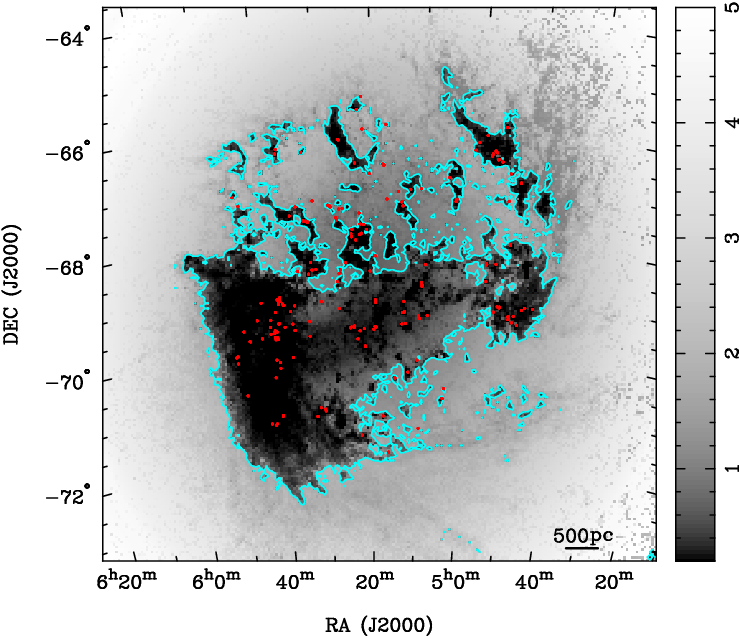}
   \caption{Map of the minimum unstable wavelength $\lambda_\mathrm{min}$,
      shown in grayscale.
      The solid contours delineate the critical boundary of $Q_{sg} = 1$, while
      the red dots mark the massive YSO candidates.
      $\lambda_\mathrm{min}$ is defined by $\lambda_\mathrm{crit}$
      (see Fig.~\ref{Fi:qsgwl}) when $Q_{sg} > 1$.}
   \label{Fi:lmin}
\end{figure}

\citet{jS04} has argued that a constant gas surface density threshold is a good
indicator in predicting the edge of the star forming disk.
He suggested that, for the LMC disk, this threshold is
$\sim$4~$M_\sun$ pc$^{-2}$.
Figure~\ref{Fi:maps}a shows the contours of 4~$M_\sun$ pc$^{-2}$ in total
neutral gas surface density.
The contours encompass the majority of the LMC disk except the central cavities
of the supergiant shells LMC-4 and LMC-8 \citep{jM80}.
Clearly, all the massive YSO candidates are located where the gas surface
density is greater than 4~$M_\sun$ pc$^{-2}$, but this constant surface density
threshold does not account for the distribution of the star forming sites,
especially in the inner parts of the LMC disk.
Therefore, Schaye's (2004) constant surface density threshold is a necessary
but not sufficient condition for star formation.
Detailed analysis of the large-scale gravitational instability, as shown in
\S~\ref{S:GI}, is still needed.

\begin{figure}
   \plotone{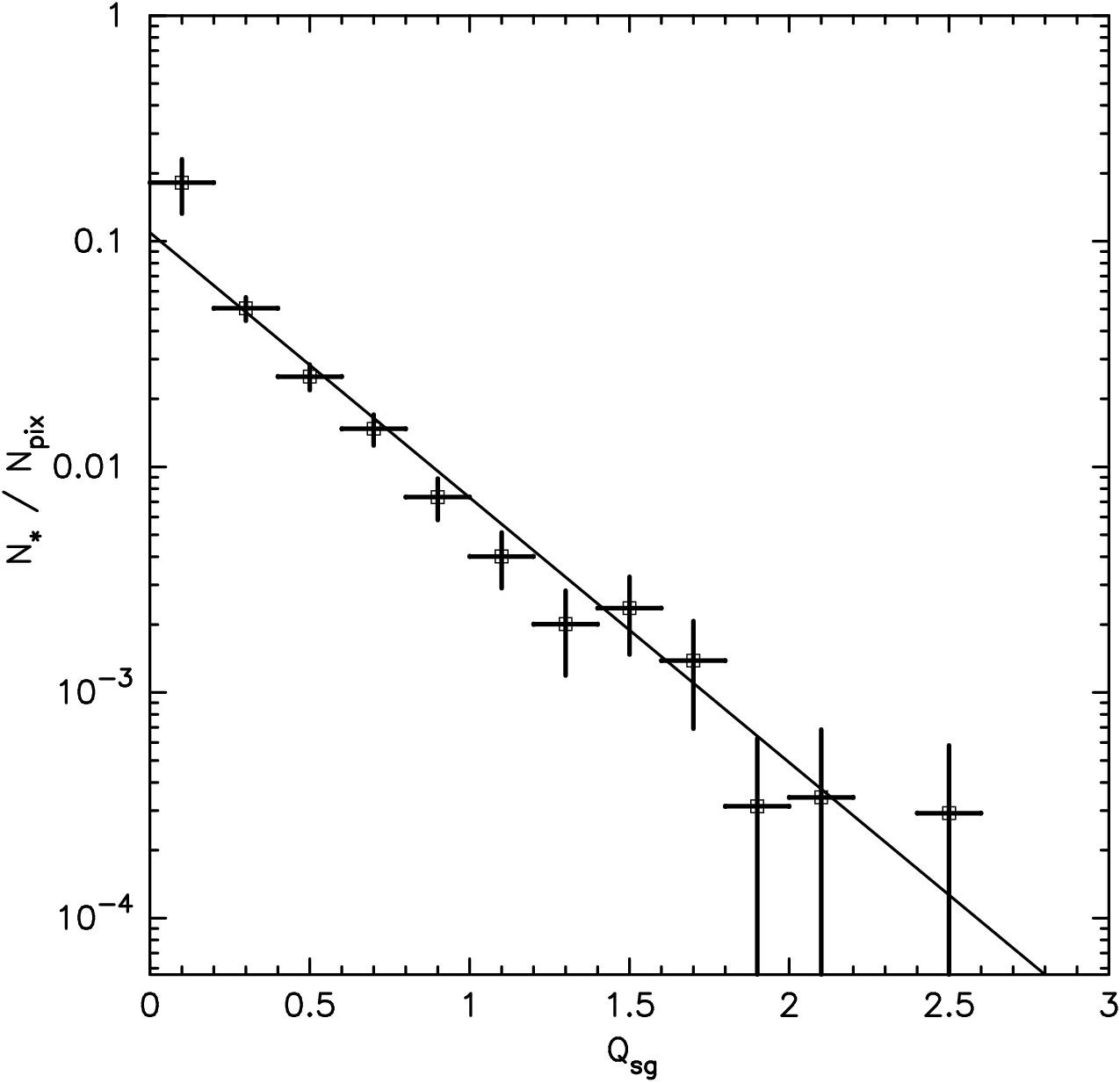}
   \caption{Ratio of the number of YSO candidates $N_*$ to the number of
      pixels $N_\mathrm{pix}$ in each $Q_{sg}$ bin, denoted by squares.
      The vertical error bars are estimated by Poisson statistics,
      while the horizontal error bars show the bin sizes.
      The \emph{solid} line is the best-fit, of slope $-2.7 \pm 0.2$.}
   \label{Fi:rate}
\end{figure}

Another measure of the response of star formation to disk instability is the
relation between $Q_{sg}$ and the effective timescale for star formation
$\tau_{sf}$.
This can be approximated if we assume the average lifetime of massive YSOs is
constant.
We normalize the number of YSOs, $N_*$, within a region having a given range of
$Q_{sg}$ by the area, as represented by the total number of pixels
$N_\mathrm{pix}$ within that region.
Figure~\ref{Fi:rate} presents $N_* / N_\mathrm{pix}$, which is $\propto
\tau_{sf}^{-1}$, for a number of $Q_{sg}$ bins.
The errors in the ratio are estimated assuming Poisson statistics on the number
of YSOs and number of pixels, while the horizontal error bars show the size of
the bins.
A clear log-linear relationship is found, implying that the response of star
formation depends exponentially on the value of $Q_{sg}$.
\citet*{LM05} also found an exponential dependence
$\tau_{sf} \propto \exp(\alpha Q_{sg})$, with $\alpha = 4.2 \pm 0.3$ from
global numerical simulations of galactic disks.
A linear fit to the data in Figure~\ref{Fi:rate} gives a slope that can be
interpreted as a local value of $\alpha = 2.7 \pm 0.2$.
This behavior appears to be a general property of nonlinear gravitational
instability.

\placefigure{Fi:rate}

We have made the simple assumption that the structure of the LMC is a
thin, flat disk with negligible stellar halo. In reality, tidal features due
to galaxy-galaxy interactions do exist in the outer parts of the disk
\citep[e.g.,][]{rV01}.  Furthermore, at some locations, there
exist multiple \ion{H}{1} velocity components, which we have simply
summed to obtain total gas surface density.
Other uncertainties stem from our assumed constant gas effective sound
speed (or velocity dispersion) and
stellar radial velocity dispersion.
These affect the relative contribution of each component to the gravitational
instability: the higher the local velocity dispersion, the less
contribution a component has. 
The sound speed of the gas we adopt, 5~km s$^{-1}$, is a typical average for
the neutral gas.
The radial velocity dispersion of stars is also uncertain.
As gravitational instability is much more sensitive to the gas component due to
its low sound speed, it is less affected by the uncertainty in the stellar
velocity dispersion.
Yet another effect to consider is the scale height of the disk, which tends to
stabilize the disk \citep{aT64,JS84}.
Given all these uncertainties, nevertheless, the exponential dependence of star
formation rate on $Q_{sg}$ that we find appears to be robust.

From our analysis of the gravitational instability of the LMC, we conclude that
the contribution of the stellar disk to gravitational instability
cannot be ignored \citep{JS84,cG92,rR01}.  Taking it into account in
the case of the LMC, we find that $\sim$85\% of the massive YSO 
candidates lie in gravitationally unstable regions, implying that star
formation occurs predominantly in these regions. It
appears that gravitational instability of the disk drives most
large-scale star formation, as proposed by \citet{bE02}, \citet{aK03}, and
\citet{LM05}.

\acknowledgments
This research was supported by NASA grants JPL1264494 and JPL1290956,
and NSF grant AST03-07854.
We would like to thank Dr.~Charles Gammie for his feedback and discussions in
this work.
We also thank the anonymous referee for the prompt review and useful
suggestions.

\clearpage

\clearpage

\clearpage

\clearpage

\clearpage

\begin{thebibliography}{}

\bibitem[Alves \& Nelson(2000)]{AN00}
Alves, D.~R., \& Nelson, C.~A.\ 2000, \apj, 542, 789

\bibitem[Blitz et al.(2007)]{BF07}
Blitz, L., Fukui, Y., Kawamura, A., Leroy, A., Mizuno, N., \&
Rosolowsky, E.\ 2007, in Protostars and Planets V,
ed. B. Reipurth, D. Jewitt, \& K. Keil (Tucson: Univ. of Arizona Press), 81
(astro-ph/0602600)

\bibitem[Dalcanton et al.(2004)Dalcanton, Yoachim, \& Bernstein]{DY04}
Dalcanton, J.~J., Yoachim, P., \& Bernstein, R.~A.\ 2004, \apj, 608, 189

\bibitem[Dib et al.(2006)Dib, Bell, \& Burkert]{DB06}
Dib, S., Bell, E., \& Burkert, A.\ 2006, \apj, 638, 797

\bibitem[Dickey et al.(1990)Dickey, Hanson, \& Helou]{DH90}
Dickey, J.~M., Hanson, M.~M., \& Helou, G.\ 1990, \apj, 352, 522 

\bibitem[Elmegreen(2002)]{bE02}
Elmegreen, B.~G.\ 2002, \apj, 577, 206

\bibitem[Feast(1999)]{mF99}
Feast, M.\ 1999, in IAU Symp.~190, New Views of the Magellanic Clouds,
ed. Y.-H. Chu, N.~B. Suntzeff, J.~E. Hesser \& D.~A. Bohlender
(San Francisco: ASP), 542

\bibitem[Freeman et al.(1983)]{FI83}
Freeman, K.~C., Illingworth, G., \& Oemler, A., Jr.\ 1983, \apj, 272, 488

\bibitem[Fukui(2007)]{yF07}
Fukui, Y.\ 2007, in IAU Symp.~237, Triggered Star Formation in a Turbulent ISM,
ed. B.~G. Elmegreen \& J. Palous (Cambridge: Cambridge Univ. Press), 31

\bibitem[Fukui et al.(2001)]{FM01}
Fukui, Y., Mizuno, N., Yamaguchi, R., Mizuno, A., \& Onishi, T.\ 2001,
\pasj, 53, L41

\bibitem[Fukui et al.(1999)]{FM99}
Fukui, Y., et al.\ 1999, \pasj, 51, 745

\bibitem[Gammie(1992)]{cG92}
Gammie, C.~F.\ 1992, Ph.D.~Thesis, Princeton University

\bibitem[Goldreich \& Lynden-Bell(1965)]{GL65}
Goldreich, P., \& Lynden-Bell, D.\ 1965, \mnras, 130, 97

\bibitem[Graff et al.(2000)]{GG00}
Graff, D.~S., Gould, A.~P., Suntzeff, N.~B., Schommer, R.~A., \&
Hardy, E.\ 2000, \apj, 540, 211

\bibitem[Jog \& Solomon(1984)]{JS84}
Jog, C.~J., \& Solomon, P.~M.\ 1984, \apj, 276, 114

\bibitem[Kennicutt(1989)]{rK89}
Kennicutt, R.~C., Jr.\ 1989, \apj, 344, 685

\bibitem[Kim et al.(1998)]{KS98}
Kim, S., Staveley-Smith, L., Dopita, M.~A., Freeman, K.~C., Sault, R.~J.,
Kesteven, M.~J., \& McConnell, D.\ 1998, \apj, 503, 674

\bibitem[Kim et al.(1999)]{KD99}
Kim, S., Dopita, M.~A., Staveley-Smith, L., \& Bessell, M.~S.\ 1999,
\aj, 118, 2797

\bibitem[Kim et al.(2003)]{KS03}
Kim, S., Staveley-Smith, L., Dopita, M.~A., Sault, R.~J., Freeman, K.~C.,
Lee, Y., \& Chu, Y.-H.\ 2003, \apjs, 148, 473

\bibitem[Kravtsov(2003)]{aK03}
Kravtsov, A.~V.\ 2003, \apjl, 590, L1

\bibitem[Kunkel et al.(1997)]{KD97}
Kunkel, W.~E., Demers, S., Irwin, M.~J., \& Albert, L.\ 1997, \apjl, 488, L129

\bibitem[Li et al.(2005)Li, Mac Low, \& Klessen]{LM05}
Li, Y., Mac Low, M.-M., \& Klessen, R.~S.\ 2005, \apj, 626, 823

\bibitem[Malhotra(1995)]{sM95}
Malhotra, S.\ 1995, \apj, 448, 138 

\bibitem[Martin \& Kennicutt(2001)]{MK01}
Martin, C.~L., \& Kennicutt, R.~C., Jr.\ 2001, \apj, 555, 301

\bibitem[Meaburn(1980)]{jM80}
Meaburn, J.\ 1980, \mnras, 192, 365

\bibitem[Mizuno et al.(1999)]{nM99}
Mizuno, N., et al.\ 1999, in Star Formation 1999, ed. T.~Nakamoto
(Nobeyama:Nobeyama Radio Observatory), 56

\bibitem[Mizuno et al.(2001)]{nM01}
Mizuno, N., et al.\ 2001, \pasj, 53, 971

\bibitem[Olsen \& Massey(2007)]{OM07}
Olsen, K.~A.~G., \& Massey, P.\ 2007, \apjl, 656, L61

\bibitem[Petric \& Rupen(2007)]{PR07}
Petric, A. O., \& Rupen, M. P.\ 2007, AJ, in press (astro-ph/0704.0279)

\bibitem[Prevot et al.(1989)]{PM89}
Prevot, L., Martin, N., \& Rousseau, J.\ 1989, \aap, 225, 303

\bibitem[Rafikov(2001)]{rR01}
Rafikov, R.~R.\ 2001, \mnras, 323, 445

\bibitem[Schaye(2004)]{jS04}
Schaye, J.\ 2004, \apj, 609, 667

\bibitem[Shostak \& van der Kruit(1984)]{SV84}
Shostak, G.~S., \& van der Kruit, P.~C.\ 1984, \aap, 132, 20 

\bibitem[Skrutskie et al.(2006)]{mS06}
Skrutskie, M.~F., et al.\ 2006, \aj, 131, 1163

\bibitem[Toomre(1964)]{aT64}
Toomre, A.\ 1964, \apj, 139, 1217

\bibitem[van der Kruit \& Shostak(1982)]{VS82}
van der Kruit, P.~C., \& Shostak, G.~S.\ 1982, \aap, 105, 351

\bibitem[van der Marel(2001)]{rV01}
van der Marel, R.~P.\ 2001, \aj, 122, 1827

\bibitem[van der Marel et al.(2002)]{VA02}
van der Marel, R.~P., Alves, D.~R., Hardy, E., \& Suntzeff, N.~B.\ 2002, \aj,
124, 2639

\bibitem[Wang \& Silk(1994)]{WS94}
Wang, B., \& Silk, J.\ 1994, \apj, 427, 759

\bibitem[Wong \& Blitz(2002)]{WB02}
Wong, T., \& Blitz, L.\ 2002, \apj, 569, 157

\end{thebibliography}
\end{document}